\documentclass[letter, 10pt, conference]{ieeeconf}  

\IEEEoverridecommandlockouts                              
\usepackage{amsmath,graphicx}
\usepackage{amssymb}
\usepackage[dvips]{epsfig}
\usepackage{mathtools}
\usepackage{subfigure}
\usepackage{multirow}

\def\mathbi#1{{#1}}

\def \pr {\mathsf{P}}

\def \s {\boldsymbol{\xi}}

\renewcommand{\boldsymbol}[1]{#1}

\newcounter{myAlgoCounter}
\usepackage{mathtools}
 \usepackage{ssCorrections}
 \xdefinecolor{colSS}{rgb}{1,0,0} 

\title{\LARGE \bf
Stability for Receding-horizon Stochastic Model Predictive Control}

\author{Joel A. Paulson$^{1,2}$, Stefan Streif$^{3}$, and Ali Mesbah$^{1,\dagger}$
\thanks{J.A.P. acknowledges funding from NSF Graduate Research Fellowship.}
\thanks{$^1$Department of Chemical and Biomolecular Engineering, University of California, Berkeley, CA 94720, USA.}
\thanks{$^{2}$Department of Chemical Engineering, Massachusetts Institute of Technology, Cambridge, MA 02139, USA.}
\thanks{$^3$Department of Computer Science and Automation, Technische Universit\"{a}t Ilmenau, 98684 Ilmenau, Germany.}
\thanks{$^{\dagger}$Corresponding author: {\tt\small mesbah@berkeley.edu}.}
}

\begin{document}

\maketitle
\thispagestyle{empty}
\pagestyle{empty}

\begin{abstract}
A stochastic model predictive control (SMPC) approach is presented for discrete-time linear systems with arbitrary time-invariant probabilistic uncertainties and additive Gaussian process noise. Closed-loop stability of the SMPC approach is established by appropriate selection of the cost function. Polynomial chaos is used for uncertainty propagation through system dynamics. The performance of the SMPC approach is demonstrated using the Van de Vusse reactions.
\end{abstract}

\section{Introduction}
\label{sec:Introduction}

Robust model predictive control (MPC) approaches have been extensively investigated over the last two decades with the goal to address control of uncertain systems with bounded uncertainties (e.g., see \cite{bem99} and the references therein). Robust MPC approaches rely on a deterministic setting and set-based uncertainty descriptions to synthesize controllers such that a worst-case objective is minimized or state constraints are satisfied with respect to all uncertainty realizations \cite{bla99}. These deterministic approaches may however lead to overly conservative control performance \cite{bem99} if the worst-case realizations have a small probability of occurrence. An approach that can alleviate the intrinsic limitations of a deterministic robust control setting is the use of stochastic descriptions of system uncertainties. This notion has led to emergence of stochastic MPC (SMPC) with chance constraints (e.g., \cite{can08,ber09,old10,cal13,mes14}), in which probabilistic descriptions of uncertainties are used to allow for prespecified levels of risk in optimal control.    

This paper investigates stability of receding-horizon SMPC. There is extensive literature that deals with tractability and stability of MPC in deterministic settings (e.g., see \cite{may00} and the references therein). However, the technical nature of arguments involved in stability of stochastic systems is significantly different, particularly in the case of unbounded uncertainties such as Gaussian process noise. In addition, there exist diverse notions of stability in the stochastic setting that are non-existent in the deterministic case \cite{cha13}. 

The work on stability of uncertain systems under receding-horizon stochastic optimal control can be broadly categorized into two research directions: first, studies that consider multiplicative process and measurement noise (e.g., \cite{can08,pri09}) and second, studies that treat process and measurement noise as additive terms in the system model (e.g., \cite{old10,ben06,hok12,far13}). The latter approaches mainly rely on the notion of affine parametrization of control inputs for finite-horizon linear quadratic problems that allows for converting the stochastic programming problem into a deterministic one. Other approaches to SMPC based on randomized algorithms have also been proposed \cite{ber09,cal13}. 

In this paper, a receding-horizon SMPC problem is presented for discrete-time linear systems with arbitrarily-shaped probabilistic time-invariant uncertainties and additive Gaussian process noise (Section~\ref{sec:ProblemStatement}). Chance constraints are incorporated into the SMPC formulation to seek tradeoffs between control performance and robustness to uncertainties. To obtain a deterministic surrogate for the SMPC formulation, the individual chance constraints are converted into deterministic expressions (in terms of the mean and variance of the stochastic system states) and a state feedback parametrization of the control law is applied (Section~\ref{sec:detSMPC}). 

This work uses the generalized polynomial chaos (gPC) framework \cite{gha91,xiu02} for probabilistic uncertainty propagation through the system dynamics to obtain a computationally tractable formulation for the presented SMPC approach (Section~\ref{sec:tractableSMPC}). The gPC framework replaces the implicit mappings (i.e., system dynamics) between the uncertain variables and the states with explicit functions in the form of a finite series of orthogonal polynomial basis functions. The Galerkin-projection method \cite{gha91} is used for analytic computation of the coefficients of the series, based on which the states' statistics are computed in a computationally efficient manner. Inspired by stability results for Markov processes \cite{mey09}, the closed-loop stability of the stochastic system is established by appropriate selection of the SMPC cost function for the unconstrained case. It is proven that the SMPC approach ensures the closed-loop stability \textsl{by design} under the corresponding receding-horizon control policy (Section~\ref{sec:stability_uncon}). The presented receding-horizon SMPC approach is used for stochastic optimal control of the Van de Vusse reactions \cite{vus64} in the presence of probabilistic parametric uncertainties, as well as additive Gaussian process noise (Section \ref{sec:CaseStudy}).

\subsection*{Notation}

Throughout this paper, $\mathbb{N} = \{1,2,\ldots\}$ is the set of natural numbers; $\mathbb{N}_0 := \mathbb{N} \cup \{0\}$; $\mathbb{R}_{+}$ is the set of nonnegative real numbers including $0$; $\mathbb{Z}_{[a,b]}:=\{a,a+1,\ldots,b\}$ is the set of integers from $a$ to $b$; $I_a$ denotes an $a\times a$ identity matrix; $\mathbf{1}_a$ denotes an $a$-dimensional vector of ones; $\mathbf{I}_a$ denotes an $a \times a$ all-ones matrix; $\mathcal{I}_A(\cdot)$ denotes the indicator function of the set $A$; $\mathbf{E}[\cdot]$ denotes the expected value; $\mathbf{E}[\cdot|x]$ denotes the conditional expected value given information $x$; $\mathbf{Var}[\cdot]$ denotes the covariance matrix; $\mathbf{Pr}[\cdot]$ denotes probability; $\mathcal{N}(\mu,\Sigma)$ denotes a Gaussian distribution with mean $\mu$ and covariance $\Sigma$; $\otimes$ denotes the Kronecker product; $\circ$ denotes the entrywise product; $\text{tr}(\cdot)$ denotes the trace of a square matrix; $\| x \|^2_A := x^\top A x$ denotes the weighted $2$-norm of $x$; and $\text{vec}(\cdot)$ denotes the column vectorization of a matrix. 

\section{Problem Statement}
\label{sec:ProblemStatement}

Consider a stochastic, discrete-time linear system 
\begin{equation} \label{e1}
x^+ = A(\theta)x + B(\theta)u + F w,
\end{equation}   
where $x \in \mathbb{R}^{n_x}$ denotes the system states at the current time instant; $x^+$ denotes the states at the next time instant; $u \in \mathbb{U}\subseteq \mathbb{R}^{n_u}$ denotes the system inputs, with $\mathbb{U}$ being a nonempty set of input constraints that is assumed to contain the origin; $\theta \in \mathbb{R}^{n_{\theta}}$ denotes the time-invariant uncertain system parameters with known finite-variance probability distribution functions (PDFs) $\pr(\theta)$; and $w \sim \mathcal{N}(0,\Sigma) \in \mathbb{R}^{n_w}$ denotes a normally distributed i.i.d. stochastic disturbance with known covariance $\Sigma \in \mathbb{R}^{n_w\times n_w}$. It is assumed that the pair $(A,B)$ is stabilizable for all uncertainty realizations $\theta$, and that the states $x$ can be observed exactly at any time. 

This work considers individual state chance constraints 
\begin{align} \label{eq:ICCs}
\mathbf{Pr}[x \in \mathcal{X}_i ] \geq \beta_i,~~i = 1,\ldots,n_{cc},
\end{align}
where $\mathcal{X}_i := \{ x \in \mathbb{R}^{n_x} | c_i^\top x \leq d_i\}$; $c_i \in \mathbb{R}^{n_x}$; $d_i \in \mathbb{R}$; $n_{cc}$ denotes the number of chance constraints; and $\beta_i \in (0, 1)$ denotes the lower bound for the probability that the $i^{th}$ state constraint should be satisfied.

This paper aims to design an SMPC approach for the stochastic system~\eqref{e1} such that the stability of the closed-loop system is guaranteed. The SMPC approach incorporates the statistical descriptions of system uncertainties into the control framework. Such a probabilistic control approach will enable shaping the states' PDFs, which is essential for seeking tradeoffs between the closed-loop performance and robustness to system uncertainties. 

Let $N \in \mathbb{N}$ denote the prediction horizon of the SMPC problem, and define $\mathbf{w}:=[w_0^\top,\ldots,w_{N-1}^\top]^\top$ as the disturbance sequence over 0 to $N-1$. A full state feedback control law $\pi$ is defined by
\begin{align}
\pi := \{ \pi_0,\pi_1(\cdot),\ldots,\pi_{N-1}(\cdot) \},
\end{align}
where $\pi_0 \in \mathbb{U}$ denotes a control action that is a function of the known current state; and $\pi_i(\cdot) : \mathbb{R}^{n_x} \longrightarrow \mathbb{U}$ denotes feedback control laws for $i=1,\ldots,N-1$. 

Let $\phi_i(x,\pi,\mathbf{w},\theta)$ denote the state predictions for the system~\eqref{e1} at time $i$ when the initial states at time $0$ are $x$, the control laws applied at times $j=0,\ldots,i-1$ are $\pi_j$, and the parameter and disturbance realizations are $w_0,\ldots,w_{i-1}$ and $\theta$, respectively. The prediction model for~\eqref{e1} is 
\begin{align} \label{eq:phipredstate}
\phi_{i+1} = A(\theta)\phi_i + B(\theta) \pi_i(\phi_i) + F w_i, \quad \phi_0 = x.
\end{align}
Note that $\{ \phi_i(x,\pi,\mathbf{w},\theta) \}_{i=0}^{N}$ represents the model predictions based on the observed states $x$. In the remainder of the paper, the explicit functional dependencies of $\phi_i$ on the initial states, control laws, and uncertainties will be dropped for notational convenience. 

A receding-horizon SMPC problem for the stochastic linear system~\eqref{e1} with time-invariant parametric uncertainties and unbounded process noise is now formulated as follows.

 \textbf{Problem 1}  \textbf{(Receding-horizon SMPC with hard input and state chance constraints):} Given the current states $x$ observed from the system \eqref{e1}, the stochastic optimal control problem is cast as 
\begin{align} \label{e_P1}
J_N^\ast(x) & := \min_{\pi} ~ J_N(x,\pi)  \\\notag
\text{s.t.:}~& \text{system model \eqref{eq:phipredstate}}, \quad \; \forall i \in \mathbb{Z}_{[0,N-1]} \\\notag
& \pi_i \in \mathbb{U}, \quad \quad \quad \quad \quad  \; \; \forall i \in \mathbb{Z}_{[0,N-1]} \\\notag
& \mathbf{Pr}[c_l^\top \phi_i \leq d_l ] \geq \beta_l, \;  \forall i \in \mathbb{Z}_{[1,N-1]}, \forall l \in \mathbb{Z}_{[1,n_{cc}]} \\\notag
& \phi_0 = x; \quad \theta \sim \pr(\theta); \quad \mathbf{w} \sim \mathcal{N}(0,\Sigma_\mathbf{w}),
\end{align}
where the objective function is defined by
\begin{align} \label{eq:JNpi}
& J_N(x,\pi) = \mathbf{E}\Big\lbrace \sum_{i=0}^{N-1} \| \phi_i \|^2_Q + \| \pi_i \|^2_R  |\Big\rbrace.
\end{align}
In~\eqref{e_P1}, $Q$ and $R$ are symmetric and positive definite weight matrices; and $\Sigma_\mathbf{w} = \text{diag}(\Sigma,\ldots,\Sigma)$. Note that the observed system states $x$ are defined as the initial states of the prediction model (i.e., $\phi_0=x$) in the SMPC program.

Solving Problem 1 is particularly challenging due to: (i) the need for parametrization of the control law $\pi$, as \eqref{e_P1} cannot be optimized over arbitrary functions $\pi$, (ii) the computational intractability of chance constraints, and (iii) the computational complexity associated with the propagation of time-invariant uncertainties through the system model~\eqref{eq:phipredstate}. In this paper, approximations are introduced to tackle the aforementioned issues in~\eqref{e_P1}. Next, a deterministic surrogate for the chance constraints is presented, followed by a state feedback parametrization for the control law $\pi$ to arrive at a deterministic formulation for Problem 1. In Section \ref{sec:tractableSMPC}, the generalized polynomial chaos framework coupled with the Galerkin projection is used for efficient propagation of the time-invariant probabilistic uncertainties $\theta$, which allows for obtaining a computationally tractable formulation for \eqref{e_P1}.

\section{Deterministic Formulation}
\label{sec:detSMPC}

\subsection{Approximation of chance constraints}
\label{sec:CC}

The following result is used to replace the chance constraints in Problem 1 with a deterministic expression in terms of the mean and variance of the predicted states.

 \textbf{Proposition 1 (Distributionally robust chance constraint \cite[Theorem 3.1]{cal06b}):} Consider an individual chance constraint of the form 
\begin{equation*} 
\mathbf{Pr} [ \mathbi{c}^\top {l} \leq 0] \geq 1-\beta, \quad \beta \in(0, 1), 
\end{equation*}
where $\mathbi{l} \in \mathbb{R}^{n_\mathbi{l}}$ denotes random quantities with known mean $\tilde{\mathbi{l}}$ and covariance $\Sigma_l$; and ${c} \in \mathbb{R}^{n_\mathbi{l}}$ denotes constants. Let $\mathcal{L}$ denote the family of all distributions with mean $\tilde{\mathbi{l}}$ and covariance $\Sigma_l$. For any $\beta \in(0, 1)$, the chance constraint
\begin{equation*} 
\inf_{\mathbi{l}  \thicksim \mathcal{L}} \; \mathbf{Pr} [ \mathbi{c}^\top {l} \leq 0] \geq 1-\beta
\end{equation*}    
(where $\mathbi{l} \thicksim \mathcal{L}$ denotes that the distribution of $\mathbi{l}$ belongs to the family $\mathcal{L}$) is equivalent to the constraint   
\begin{equation*} 
\mathbf{E}[\mathbi{c}^\top {l}] + \kappa_{\beta}\sqrt{\mathbf{Var}[\mathbi{c}^\top {l}]} \leq 0, \quad \kappa_{\beta}=\sqrt{(1-\beta)/\beta}, 
\end{equation*}  
where $\mathbf{E}[\mathbi{c}^\top {l}]=c^\top\tilde{\mathbi{l}}$; and $\mathbf{Var}[\mathbi{c}^\top {l}] = {c}^\top \Sigma_l {c}$. $\hfill \blacksquare$ 

Using Proposition 1, the chance constraints in \eqref{e_P1} can be replaced with the deterministic expression 
\begin{align} \label{e_cc3}
c_l^\top \mathbf{E}[\phi_i] + \kappa_{1-\beta_l} \sqrt{c_l^\top \textbf{Var}[\phi_i] c_l} \leq d_l,
\end{align}
which guarantees that the state constraint $\phi_i \in \mathcal{X}_l$ is satisfied with at least probability $\beta_l$. 

\subsection{State feedback parametrization of control law}

To incorporate feedback control over the prediction horizon, the control law $\pi$ is parametrized as an affine function of the states
\begin{align} \label{eq:sf_policy}
\pi_i(\phi_i) := g_i + L_i \phi_i, \quad \forall i \in \mathbb{Z}_{[0,N-1]},
\end{align}
where $g_i \in \mathbb{R}^{n_u}$ and $L_i \in \mathbb{R}^{n_u \times n_x}$ are affine terms and feedback gains, respectively. 

Let $\tilde{L} = \{ L_0,\ldots,L_{N-1} \}$ and $\tilde{g} = \{ g_0,\ldots,g_{N-1} \}$ denote the set of decision variables in~\eqref{e_P1} to be optimized over the prediction horizon $N$. A deterministic reformulation of Problem 1 using the control law~\eqref{eq:sf_policy} is stated as follows.

\textbf{Problem 2} \textbf{(Deterministic formulation for SMPC with hard input and state chance constraints):}
\begin{align} \label{e_P2}
& ~ \underset{(\tilde{L},\tilde{g})}{\min} ~ J_N(x,\tilde{L},\tilde{g}) \\\notag
& \begin{array}{lll}
\text{s.t.:} & \phi_{i+1} = A(\theta)\phi_i + B(\theta) \pi_i + w_i, & \forall i\in\mathbb{Z}_{[0,N-1]} \\
& \pi_i = g_i + L_i \phi_i \in \mathbb{U}, & \forall i \in \mathbb{Z}_{[0,N-1]} \\
& c_l^\top \mathbf{E}[\phi_i] + \kappa_{1-\beta_l} \sqrt{c_l^\top \textbf{Var}[\phi_i] c_l} \leq d_l, & \forall i \in \mathbb{Z}_{[1,N-1]} \\
& & \forall l \in \mathbb{Z}_{[1,n_{cc}]} \\
& \phi_0 = x; \; \; \theta \sim \pr(\theta); \; \;	\mathbf{w} \sim \mathcal{N}(0,\Sigma_\mathbf{w}).
\end{array}
\end{align}

In Problem 2, the objective function and chance constraints are defined in terms of the mean and variance of the predicted states $\phi_i$. Next, the gPC framework is used to propagate uncertainties $\theta$ and $\mathbf{w}$ through the system model~\eqref{eq:phipredstate}. This will enable approximating the moments of $\phi_i$ to solve \eqref{e_P2}.

\textbf{Remark 1}: In general, it is impossible to guarantee input constraint satisfaction for a state feedback control law in the presence of unbounded disturbances unless $\{L_i = 0\}_{i=0}^{N-1}$ (i.e., \eqref{eq:sf_policy} takes the form an open-loop control law). Hence, the hard input constraints are not considered in the remainder of this paper by assuming $\mathbb{U} = \mathbb{R}^{n_u}$.\footnote{Alternative approaches are truncating the distribution of the stochastic disturbances to robustly guarantee the input constraints over a bounded set, or defining input chance constraints (e.g., see \cite{far13}).}       


\section{Tractable Stochastic Predictive Control}
\label{sec:tractableSMPC}

\subsection{Polynomial chaos for uncertainty propagation}
\label{sec:PCE}

The gPC framework enables approximation of a stochastic variable $\psi(\s)$ in terms of a finite series expansion of orthogonal polynomial basis functions
\begin{align} \label{e_PCE}
\psi(\s) \approx \hat{\psi}(\s) \coloneqq  \sum\limits_{k=0}^{p}a_k\varphi_{k}(\s) = \mathbf{a}^\top\Lambda(\s),  
\end{align}
where $\mathbf{a} \coloneqq [a_0, \ldots, a_p]^{\top}$ denotes the vector of expansion coefficients; $\Lambda(\s) \coloneqq [\varphi_{0}(\s), \ldots, \varphi_{p}(\s)]^\top$ denotes the vector of basis functions $\varphi_{k}$ of maximum degree $m$ with respect to the random variables $\s$; and $p+1=\frac{(n_\xi+m)!}{n_\xi!m!}$ denotes the total number of terms in the expansion. The basis functions belong to the Askey scheme of polynomials, which encompasses a set of orthogonal basis functions in the Hilbert space defined on the support of the random variables \cite{xiu02}. This implies that $\langle \varphi_i(\s), \varphi_j(\s) \rangle = \langle \varphi_i^2(\s) \rangle\delta_{ij}$, where $\langle h(\s), g(\s) \rangle=\int_{\Omega}h(\s)g(\s)\pr({\s})d\s$ denotes the inner product induced by $\pr({\s})$, and $\delta_{ij}$ denotes the Kronecker delta function. Hence, the coefficients $a_k$ in~\eqref{e_PCE} are defined by $a_k = \frac{\langle \hat{\psi}(\s) , \varphi_{k}(\s) \rangle}{\langle \varphi_{k}(\s),\varphi_{k}(\s) \rangle}$. For the linear and polynomial systems, the integrals in the inner products can be computed analytically \cite{gha91}. The basis functions $\varphi_{k}$ are chosen in accordance with the PDFs of $\s$.

\subsection{Evaluation of multivariate PDF of states}

The time evolution of the multivariate PDF of states (given $\phi_0=x$) describes the propagation of stochastic uncertainties $\theta$ and $\{w_i\}_{i=0}^{N-1}$ through the system model \eqref{eq:phipredstate}. For a particular realization of $\{w_i\}_{i=0}^{N-1}$, the propagation of $\theta$ through \eqref{eq:phipredstate} can be efficiently described using the gPC framework. The propagation of $\theta$ will result in the conditional predicted states' PDF $\pr(\phi_{i+1}|\{w_s\}_{s=0}^i)$, which can be integrated over all possible realizations of $\{w_s\}_{s=0}^i$ to obtain the complete PDF $\pr(\phi_{i+1})$ at every time, i.e.,
\begin{align} \label{e_pdf}
\pr(\phi_{i+1}) & = \int_{-\infty}^{\infty} \pr(\phi_{i+1}|\{w_s\}_{s=0}^i)\prod_{s=0}^i\pr(w_s)dw_s.
\end{align}
Since $\pr(w_s)$ is a Gaussian distribution, \eqref{e_pdf} simplifies substantially when evaluating the moments of $\pr(\phi_{i+1})$.

To use the gPC framework, each element of the predicted states $\phi$ and control laws $\pi$, as well as the system matrices $A(\theta)$ and $B(\theta)$ in \eqref{eq:phipredstate} are approximated with a finite PC expansion of the form \eqref{e_PCE}. Define $\Phi_{i,t} = [a_{i_0,t},\ldots,a_{i_p,t}]^\top$ and $\Pi_{i,t} = [b_{i_0,t},\ldots,b_{i_p,t}]^\top$ to be the set of PC expansion coefficients for the $i^{th}$ predicted states and inputs at time $t$, respectively. Concatenate the latter two vectors into vectors $\mathbf{\Phi}_t := [\Phi_{1,t}^\top,\ldots,\Phi_{n_x,t}^\top ]^\top \in \mathbb{R}^{n_x(p+1)}$ and $\mathbf{\Pi}_t := [\Pi_{1,t}^\top,\ldots,\Pi_{n_u,t}^\top ]^\top \in \mathbb{R}^{n_u(p+1)}$. The Galerkin projection \cite{pau14} can now be used to project the error in the truncated expansion approximation of \eqref{eq:phipredstate} onto the space of orthogonal basis functions $\{ \varphi_{k}\}_{k=0}^p$, yielding
\begin{align} \label{eq:Xdet}
\mathbf{\Phi}_{i+1} = \mathbf{A} \mathbf{\Phi}_i + \mathbf{B} \mathbf{\Pi}_i + \mathbf{F} w_i,
\end{align}
where
\begin{align*}
\mathbf{A} = \sum_{k=0}^p A_k \otimes \Psi_k;~\mathbf{B} = \sum_{k=0}^p B_k \otimes \Psi_k;~\mathbf{F} = F \otimes e_{p+1};
\end{align*}
\begin{align*}
\Psi_k := \begin{bmatrix}
\sigma_{0k0} &\cdots &\sigma_{0kp} \\
\vdots &\ddots &\vdots \\
\sigma_{pk0} &\cdots &\sigma_{pkp}
\end{bmatrix};
\end{align*}
$A_k$ and $B_k$ are the projections of $A(\theta)$ and $B(\theta)$ onto the $k^{th}$ basis function $\varphi_k$; $\sigma_{ijk} = \langle \varphi_i, \varphi_j, \varphi_k \rangle / \langle \varphi_i^2 \rangle$; and $e_a = [1,0,\ldots,0]^\top \in \mathbb{R}^{a}$. The orthogonality property of the multivariate polynomials in the PC expansions is exploited to efficiently compute the moments of the conditional PDF $P(\phi_{t+1}|\{w_s\}_{s=0}^t)$ using the coefficients $\mathbf{\Phi}_{t+1}$. The first two conditional moments of the $i^{th}$ predicted states are defined by
\begin{subequations} \label{eq:conditionalmoment}
\begin{align}
 \mathbf{E}[\phi_{i,t+1} | \{w_s\}_{s=0}^t] & \approx a_{i_0,t+1}(w_0,\ldots,w_t) \\
 \mathbf{E}\big[\phi_{i,t+1}^2 | \{w_s\}_{s=0}^t \big] & \approx \sum_{k=0}^p  a_{i_k,t+1}^2(w_0,\ldots,w_t) \langle \varphi_k^2 \rangle.
\end{align}
\end{subequations}

Similarly, the state feedback control law \eqref{eq:sf_policy} is projected
\begin{align} \label{eq:Udet}
\mathbf{\Pi}_i = \mathbf{g}_i + \mathbf{L}_i \mathbf{\Phi}_i,
\end{align}
where $\mathbf{g}_i = g_i \otimes e_{p+1}$ and $\mathbf{L}_i = L_i \otimes I_{p+1}$. Since $\{ w_i \}_{i=0}^{N-1}$ is assumed to be Gaussian white noise, $\{ \mathbf{\Phi}_i \}_{i=1}^{N}$ is a Gaussian process with mean $\bar{\mathbf{\Phi}}_i$ and covariance $\Gamma_i$ defined by
\begin{subequations} \label{eq:gPCprop}
\begin{align}
 \bar{\mathbf{\Phi}}_{i+1} & = (\mathbf{A} + \mathbf{B}\mathbf{L}_i) \bar{\mathbf{\Phi}}_i + \mathbf{B}\mathbf{g}_i \\
 \Gamma_{i+1} & = (\mathbf{A} + \mathbf{B}\mathbf{L}_i) \Gamma_i (\mathbf{A} + \mathbf{B}\mathbf{L}_i)^\top + \mathbf{F} \Sigma \mathbf{F}^\top. 
\end{align}
\end{subequations}
Note that $(\bar{\mathbf{\Phi}}_i,\Gamma_i)$ is initialized using the current states $\phi_0 = x$ via projection (i.e., $\bar{\mathbf{\Phi}}_0 = x \otimes e_{p+1}$ and $\Gamma_0 = 0$). Using \eqref{eq:conditionalmoment}, \eqref{eq:gPCprop}, and the law of iterated expectation, tractable expressions for describing the first two moments of $\pr(\phi_{i,t+1})$ are derived as
\begin{align} \label{eq:state_mean}
\mathbf{E}[\phi_{i,t+1}] & = \mathbf{E}\big[\mathbf{E}[\phi_{i,t+1}|\{w_s\}_{s=0}^t]\big] \\\notag
& \approx \mathbf{E}[a_{i_0,t+1}(w_0,\ldots,w_t)] = \bar{a}_{i_0,t+1}
\end{align}
\begin{align} \label{eq:state_variance}
& \mathbf{E}[\phi_{i,t+1}^2] = \mathbf{E}\big[\mathbf{E}[\phi^2_{i,t+1}|\{w_s\}_{s=0}^t]\big] \\\notag
& \approx \mathbf{E}\big[ \sum_{k=0}^p  a_{i_k,t+1}^2(w_0,\ldots,w_t) \langle \varphi_k^2 \rangle \big] = \sum_{k=0}^p \mathbf{E}[ a_{i_k,t+1}^2] \langle \varphi_k^2 \rangle \\\notag
& = \sum_{k=0}^p \big[ \bar{a}^2_{i_k,t+1} + \Gamma_{i_ki_k,t+1} \big] \langle \varphi_k^2 \rangle. 
\end{align}

\subsection{Tractable SMPC formulation using gPC}
\label{sec:ReformulationOfProblem1}

In this section, the goal is to use the gPC framework to obtain a tractable approximation of Problem 2 in terms of $(\bar{\mathbf{\Phi}}_i,\Gamma_i)$. The objective function~\eqref{eq:JNpi} is rewritten exactly as  
\begin{align*}
J_N  = \mathbf{E}\bigg\lbrace \sum_{i=0}^{N-1} & \mathbf{E}\Big[\| \phi_i \|^2_Q \big| \{ w_s \}_{s=0}^{i-1}\Big] + \mathbf{E}\Big[\| \pi_i \|_R^2 \big| \{ w_s \}_{s=0}^{i-1}\Big] \bigg\rbrace,
\end{align*}
where the conditional moments can be approximated by \eqref{eq:conditionalmoment}
\begin{align} \label{eq:of_1}
J_N \approx V_N := \mathbf{E}\Big\lbrace \sum_{i=0}^{N-1} \| \mathbf{\Phi}_i \|^2_\mathbf{Q} + \| \mathbf{\Pi}_i \|^2_\mathbf{R} \Big\rbrace.
\end{align}
In~\eqref{eq:of_1}, $\mathbf{Q} = Q \otimes W$, $\mathbf{R} = R \otimes W$, and $W = \text{diag}(\langle \varphi_0^2 \rangle,\langle \varphi_1^2 \rangle,\ldots,\langle\varphi_{p}^2\rangle)$. Substituting the control law \eqref{eq:Udet} in~\eqref{eq:of_1} and rearranging the resulting equation yields
\begin{align*}
V_N = \sum_{i = 0}^{N-1} & \| \bar{\mathbf{\Phi}}_i \|^2_{\mathbf{Q} + \mathbf{L}_i^\top \mathbf{R} \mathbf{L}_i} + \text{tr}\big\{ (\mathbf{Q}+ \mathbf{L}_i^\top \mathbf{R} \mathbf{L}_i) \Gamma_i \big\} +\\[-2mm]\notag
& \| \mathbf{g}_i \|^2_{\mathbf{R}} + 2 \mathbf{g}_i^\top \mathbf{R} \mathbf{L}_i \Omega^\top \bar{\mathbf{\Phi}}_i + \| \bar{\mathbf{\Phi}}_N \|^2_{\mathbf{S}} + \text{tr}\big\{\mathbf{S} \Gamma_N \big\},
\end{align*}
where $\Omega = (I_{n_x} \otimes e_{p+1})$.

To compute the chance constraints, \eqref{eq:state_mean} and \eqref{eq:state_variance} are used to approximate the mean and variance in \eqref{e_cc3} by 
\begin{align} \label{eq:Exp}
\mathbf{E}[\phi_i] \approx \Omega^\top \bar{\mathbf{\Phi}}_i
\end{align}
and
\begin{align} \label{eq:Var}
& \mathbf{E}[\phi_i \phi_i^\top] \approx \Big[ \{ (I_{n_x} \otimes \mathbf{1}_{p+1}) \circ (\bar{\mathbf{\Phi}}_i \mathbf{1}_{n_x}^\top) \}^\top (\mathbf{I}_{n_x} \otimes W) \\\notag
&  \{ (I_{n_x} \otimes \mathbf{1}_{p+1}) \circ (\bar{\mathbf{\Phi}}_i \mathbf{1}_{n_x}^\top) \} \Big] + M(I_{n_x}\otimes\text{vec}(\Gamma_i)),
\end{align}
respectively, where 
\begin{align*}
M = \begin{bmatrix}
\text{vec}(E_{1,1} \otimes W)^\top &\cdots &\text{vec}(E_{1,n_x} \otimes W)^\top \\
\vdots &\ddots &\vdots \\
\text{vec}(E_{n_x,1} \otimes W)^\top &\cdots &\text{vec}(E_{n_x,n_x} \otimes W)^\top
\end{bmatrix};
\end{align*}
and $E_{i,j} \in \mathbb{R}^{n_x \times n_x}$ is a binary matrix with a value $1$ in only the $(i,j)^{th}$ entry. The tractable formulation of Problem 2 can now be presented as follows.


\textbf{Problem 3} \textbf{(Tractable formulation for SMPC with state chance constraints):}
\begin{align} \label{e_P3}
& \min_{(\tilde{L},\tilde{g})} ~ V_N(x,\tilde{L},\tilde{g}) \\[1mm]\notag
& \begin{array}{lll}
\text{s.t.:} & \bar{\mathbf{\Phi}}_{i+1} = (\mathbf{A} + \mathbf{B}\mathbf{L}_i) \bar{\mathbf{\Phi}}_i + \mathbf{B}\mathbf{g}_i, & \forall i \in \mathbb{Z}_{[0,N-1]} \\[1mm]
& \Gamma_{i+1} = (\mathbf{A} + \mathbf{B}\mathbf{L}_i) \Gamma_i (\mathbf{A} + \mathbf{B}\mathbf{L}_i)^\top &  \\
& ~~~~~~~~~~ + \mathbf{F} \Sigma \mathbf{F}^\top, & \forall i \in \mathbb{Z}_{[0,N-1]} \\[1mm]
& \multicolumn{2}{l}{f^{[l]}_{cc}(\bar{\mathbf{\Phi}}_{i},\Gamma_{i}) \leq 0,~~~~~~~~\forall i \in \mathbb{Z}_{[1,N-1]},~ \forall l \in \mathbb{Z}_{[1,n_{cc}]}} \\[1mm]
& \bar{\mathbf{\Phi}}_0 = x \otimes e_{p+1}; \quad \Gamma_0 = 0, 
\end{array}
\end{align}
where $f^{[l]}_{cc}$ denotes the $l^{th}$ chance constraint as a function of $(\bar{\mathbf{\Phi}}_i,\Gamma_i)$ that is straightforwardly derived by substituting \eqref{eq:Exp} and \eqref{eq:Var} into the deterministic chance constraint \eqref{e_cc3}.

\section{Stability Analysis for Unconstrained Stochastic Model Predictive Control}
\label{sec:stability_uncon}

The initialization strategy (i.e., $\bar{\mathbf{\Phi}}_0 = x \otimes e_{p+1},\Gamma_0 = 0$) used for solving Problem 3 relies on the current state observations $x$. This implies that the SMPC problem is solved conditioned on the measured states $x$. However, unbounded disturbances $w$ act on the closed-loop states, which makes it impossible to assert convergence of the states to any compact set under any control policy.\footnote{There will almost surely be excursions of the states beyond any compact set infinitely often over an infinite time horizon \cite{cha13}.} The fact that the states $x$ can jump anywhere in $\mathbb{R}^{n_x}$ also makes it difficult to guarantee feasibility of chance constraints. An approach for guaranteeing feasibility is to switch the initialization strategy such that it corresponds to the open-loop control problem while adding appropriate terminal constraints \cite{far13}. To obviate the use of such approaches, the stability of the SMPC approach presented in Problem 3 is established for the unconstrained case.  

Consider discrete-time Markov processes $\{x_t\}_{t\in\mathbb{N}_0}$, where the PDF of the states $x_{t+1}$ is conditionally independent of the past states $\{x_s\}_{s=0}^{t-1}$ given the present states $x_t$. The stability under study concerns boundedness of sequences of the form $\{\mathbf{E}[h(x_t)|x_0=x]\}_{t\in\mathbb{N}_0}$, where $h$ is some norm-like function \cite{cha13}. The theory of stability for discrete-time Markov processes entails a \textit{negative drift condition} \cite{mey09}.

\textbf{Proposition 2 (Geometric Drift \cite{cha13}):} Let $\{x_t\}_{t \in \mathbb{N}_0}$ denote a Markov process. Suppose there exists a measurable function $V : \mathbb{R}^{n_x} \longrightarrow \mathbb{R}_{+}$, a compact set $D \subset \mathbb{R}^{n_x}$ such that $\mathbf{E}[V(x_1)|x_0=x] $, $\forall x \not\in D$, $\sup_{x\in D}\mathbf{E}[V(x_1)|x_0=x] = b$ for some constants $b \geq 0$, and $\lambda \in [0,1)$. Then, $\mathbb{E}[V(x_t) | x_0 = x] \leq \lambda^t V(x) + b(1-\lambda)^{-1}, ~ \forall x \in \mathbb{R}^{n_x}, ~\forall t \in \mathbb{N}_0$. This implies that the sequence $\{\mathbf{E}[V(x_t)|x_0=x]\}_{t\in\mathbb{N}_0}$ is bounded $\forall x \in \mathbb{R}^{n_x}$. A \textit{geometric drift condition} is also satisfied for states outside a compact set $D$, i.e.,
\begin{align*}
\mathbf{E}[V(x_1)|x_0=x] - V(x) \leq -(1-\lambda) V(x),~\forall x\not\in D. \quad \; \; \blacksquare
\end{align*}

In what follows, the stability results for stochastic predictive control \cite{cha13} are extended to deal with probabilistic parametric uncertainties. This is done by appropriate selection of the cost function such that a drift condition on the optimal value function can be established.

\subsection{Preliminaries}

Let $n:= n_x(p+1)$ and $r := n_u(p+1)$ denote the dimensions of the gPC projected states and inputs, respectively. A terminal cost $\| \mathbf{\Phi}_N \|_P^2$ is included into the objective function~\eqref{eq:of_1}, where $P = P^\top > 0$ is the solution to the Lyapunov equation
\begin{align} \label{eq:Lyap}
(\mathbf{A}+\mathbf{B}\mathbf{K})^\top P & (\mathbf{A}+\mathbf{B}\mathbf{K}) - P  = -(1+\delta)\mathbf{M}
\end{align}
with $\delta > 0$; $\mathbf{M} := \mathbf{Q}+\mathbf{K}^\top \mathbf{R}\mathbf{K}$; and $\mathbf{K} := K \otimes I_{p+1}$. The objective function $V_N$ is now stated as
\begin{align} \label{eq:VN}
& V_N(\mathbf{\Phi},\tilde{L},\tilde{g}) \\\notag
& = \mathbf{E}\Big\lbrace \sum_{i=0}^{N-1} \| \mathbf{\Phi}_i \|^2_\mathbf{Q} + \| \mathbf{\Pi}_i \|^2_\mathbf{R} + \| \mathbf{\Phi}_N \|^2_P \big| \mathbf{\Phi}_0 = \mathbf{\Phi} \Big\rbrace.
\end{align}
Since the pair $(A(\theta),B(\theta))$ is assumed to be stabilizable for all realizations of $\theta$, there exists a feedback gain $K$ and $P > 0$ that satisfies \eqref{eq:Lyap} \cite{fis09}. Note that $\mathbf{\Phi} \in \mathbb{R}^{n}$ represents the initial conditions in the PC expansion coefficient space. In \eqref{eq:VN}, the stage cost $c : \mathbb{R}^{n}\times \mathbb{R}^{r} \longrightarrow \mathbb{R}_{+}$ and the final cost $c_f : \mathbb{R}^{n} \longrightarrow \mathbb{R}_{+}$ are denoted by
\begin{align} \label{eq:stagecost}
 c(\mathbf{\Phi},\mathbf{\Pi}) &= \| \mathbf{\Phi} \|_\mathbf{Q}^2 + \| \mathbf{\Pi} \|_\mathbf{R}^2 \\\label{eq:finalcost}
 c_f(\mathbf{\Phi}) &= \| \mathbf{\Phi} \|^2_P.
\end{align}

\textbf{Problem 4} \textbf{(Unconstrained SMPC):} For any initial condition $\mathbf{\Phi} \in \mathbb{R}^{n}$, the unconstrained $N$-horizon stochastic optimal control problem is stated as
\begin{align} \label{eq:VNuncon}
& \min_{(\tilde{L},\tilde{g})}~V_N(\mathbf{\Phi},\tilde{L},\tilde{g}) \\[1mm]\notag
& \begin{array}{lll}
\text{s.t.:~\eqref{eq:Xdet} and \eqref{eq:Udet}}, \;  \forall i \in \mathbb{Z}_{[0,N-1]}. 
\end{array}
\end{align}

Let $\pi^\ast$ denote the optimal state feedback policy computed from \eqref{eq:VNuncon} with parametrization \eqref{eq:sf_policy} (i.e., $\pi_i^\star(x) = g_i^\star(x) + L_i^\star(x) x, ~ \forall i \in \mathbb{Z}_{[0,N-1]}$). Given the state $x_t$ at time $t$, Problem 4 is implemented in a receding-horizon mode that entails: (i) solving \eqref{eq:VNuncon} for $\pi^\star$ with $\mathbf{\Phi} = x_t \otimes e_{p+1}$, (ii) applying the first element $\pi_0^\star$ to the system~\eqref{e1}, and (iii) shifting to time $t+1$ and repeating the preceding steps.

As the true system \eqref{e1} has fixed parameter values ($\theta = \hat{\theta}$), it evolves as a Markov process. Under the policy $\{ \pi_0^\star,\pi_0^\star,\ldots \}$, the system \eqref{e1} generates a state trajectory $\{x_t\}_{t\in\mathbb{N}_0}$ via the recursion
\begin{align} \label{eq:cl_state_rec}
x_{t+1} = \hat{A}x_t + \hat{B}\pi^\star_0(x_t) + Fw_t,~x_0 \text{ given},~ \forall t \in \mathbb{N}_0,
\end{align}
where $\hat{A} = A(\hat{\theta})$ and $\hat{B} = B(\hat{\theta})$ are the true plant matrices.

\subsection{Stability through boundedness of value function}

Let $(\tilde{L}^\star,\tilde{g}^\star)$ denote the optimal control parameters obtained by solving \eqref{eq:VNuncon} for a given initial condition. Denote the optimal value function by $V_N^\star(\mathbf{\Phi}):=V_N(\mathbf{\Phi},\tilde{L}^\star,\tilde{g}^\star)$. The following two propositions form the basis for the main stability result for Problem~4, which is presented in Theorem~1. 

\textbf{Proposition 3:} The stage cost \eqref{eq:stagecost}, final cost \eqref{eq:finalcost}, and controller $\mathbf{K}\mathbf{\Phi}$ satisfy
\begin{subequations}
\begin{align}
& \sup_{\mathbf{\Phi} \in D} \Big\{ c(\mathbf{\Phi},\mathbf{K}\mathbf{\Phi}) - c_f(\mathbf{\Phi}) \label{eq:Lem3_PhiinD}\\[-4mm]\notag
& ~~~~~~~ + \mathbf{E}\big[c_f\big((\mathbf{A}+\mathbf{B}\mathbf{K})\mathbf{\Phi}+\mathbf{F}w_0\big) | \mathbf{\Phi}\big] \Big\} \leq b \\
& c(\mathbf{\Phi},\mathbf{K}\mathbf{\Phi}) - c_f(\mathbf{\Phi}) \label{eq:Lem3_PhinotinD}\\\notag 
& ~ + \mathbf{E}\big[c_f\big((\mathbf{A}+\mathbf{B}\mathbf{K})\mathbf{\Phi}+\mathbf{F}w_0\big)| \mathbf{\Phi}\big] \leq 0,~\forall \mathbf{\Phi} \not\in D
\end{align}
\end{subequations}
for some constant $b \geq 0$ and a bounded measurable set $D := \big\{ z \in \mathbb{R}^{n} | z^\top \mathbf{M} z \leq \frac{1}{\delta}\text{tr}(\mathbf{F}^\top P \mathbf{F} \Sigma ) \big\}$.

\noindent Proof: See Appendix A. $\hfill$ $\blacksquare$

\textbf{Proposition 4:} For all $\mathbf{\Phi} \in \mathbb{R}^{n}$, the optimal value function satisfies $V_N^\star(\mathbf{\Phi}) \leq c_f(\mathbf{\Phi}) + Nb$.

\noindent Proof: See Appendix B. $\hfill$ $\blacksquare$

Recall that $x_1$ is computed from \eqref{eq:cl_state_rec}. It is assumed that 
\begin{align} \label{eq:assumption}
\mathbf{E}\big[ V_N^f(x_1 & \otimes e_{p+1}) | x_0 = x  \big] \\\notag
& \leq \mathbf{E}\big[ V_N^f(\mathbf{\Phi}^\star_{1}) | \mathbf{\Phi}^\star_0 = x \otimes e_{p+1} \big],~\forall x \in \mathbb{R}^{n_x}.
\end{align}

\textbf{Theorem 1:} Consider the system \eqref{e1} at a fixed $\theta = \hat{\theta}$, and the stochastic optimal control problem \eqref{eq:VNuncon}. Suppose that assumption \eqref{eq:assumption} holds. Then, $\{ \mathbf{E}[V_N^\star(x_t \otimes e_{p+1})| x_0 = x] \}_{t \in \mathbb{N}_0}$ is bounded for each $x \in \mathbb{R}^{n_x}$.

\noindent Proof: See Appendix C. $\hfill$ $\blacksquare$

As shown in the proof of Theorem~1, $V_N^\star$ satisfies a geometric drift condition outside of some compact set of $\mathbb{R}^{n_x}$. Hence, the receding-horizon SMPC approach in Problem 4 results in a bounded objective for all times such that the discrete-time Markov system is stochastically stable.

\section{Case Study: Van de Vusse Reactor}
\label{sec:CaseStudy}

The Van de Vusse series of reactions in an isothermal continuous stirred-tank reactor \cite{vus64} is considered to evaluate the performance of the receding-horizon SMPC approach (i.e., Problem 3). The dynamic evolution of the concentration of compounds $A$ and $B$ (denoted by $C_A$ and $C_B$, respectively) is described by
\begin{align} \label{e_r}
\begin{array}{ll}
\dot{C}_A &= -k_1C_A - k_3C_A^2 - C_Au \\
\dot{C}_B &= k_1C_A - k_2C_B - C_Bu,   
\end{array}
\end{align}    
where $k_1$, $k_2$, and $k_3$ denote the rate constants; and $u$ is the dilution rate. Linearizing~\eqref{e_r} around an operating point and discretizing the linearized model with a sampling time of $0.002$ \cite{sco98} results in a system of the form~\eqref{e1} with 
\begin{align*}  
A = \left( \begin{array}{cc}
\theta_1 & 0  \\
0.088 & 0.819  \end{array} \right) \quad \quad B = \left( \begin{array}{c}
-0.005  \\
-0.002  \end{array} \right),
\end{align*}
where $\theta_1$ has the four-parameter beta distribution $\beta(0.923,0.963,2,5)$. The noise matrix in~\eqref{e1} is assumed to be identity and $\Sigma = 10^{-4}I_2$. The states of the linearized model are defined in terms of the deviation variables $x_1$ and $x_2$. The initial states have PDFs $x_1 \sim \mathcal{N}(0.5,0.01)$ and $x_2 \sim \mathcal{N}(0.1,0.01)$, respectively. The control objective is to keep the states at the desired operating point in the presence of probabilistic uncertainties and process noise. In addition, $x_2$ should remain below the limit $0.17$ (i.e., $x_2 < 0.17$).    

To formulate the SMPC problem in~\eqref{e_P3}, a fifth-order expansion of Jacobi polynomials is used to propagate the time-invariant uncertainties. The weight matrix $Q$ in the objective function~\eqref{eq:JNpi} is defined as $I_2$ (while $R=0$), implying that there is equal importance for both states to have minimum variance around the operating point. The probability $\beta$ in the chance constraint imposed on $x_2$ is $0.95$.

\begin{figure}[t!] 
\centering
\includegraphics[width=225pt]{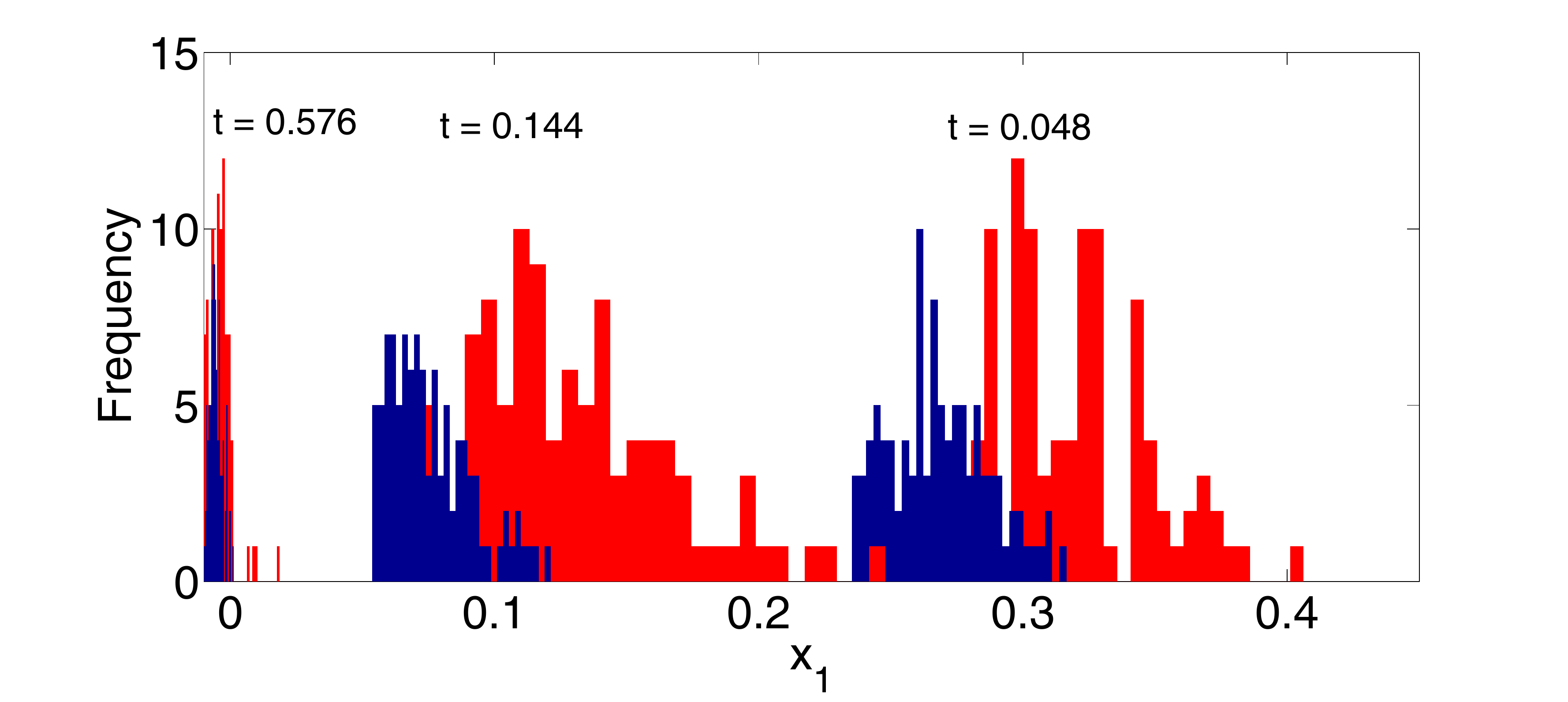}
\caption{Histograms of $x_1$ at different times obtained from $100$ closed-loop simulations of SMPC (blue) and nominal MPC (red). SMPC leads to smaller mean and variance of $x_1$.}
\label{fig1}
\end{figure}

\begin{figure}[t!] 
\centering
\subfigure[SMPC]{
\includegraphics[width=225pt]{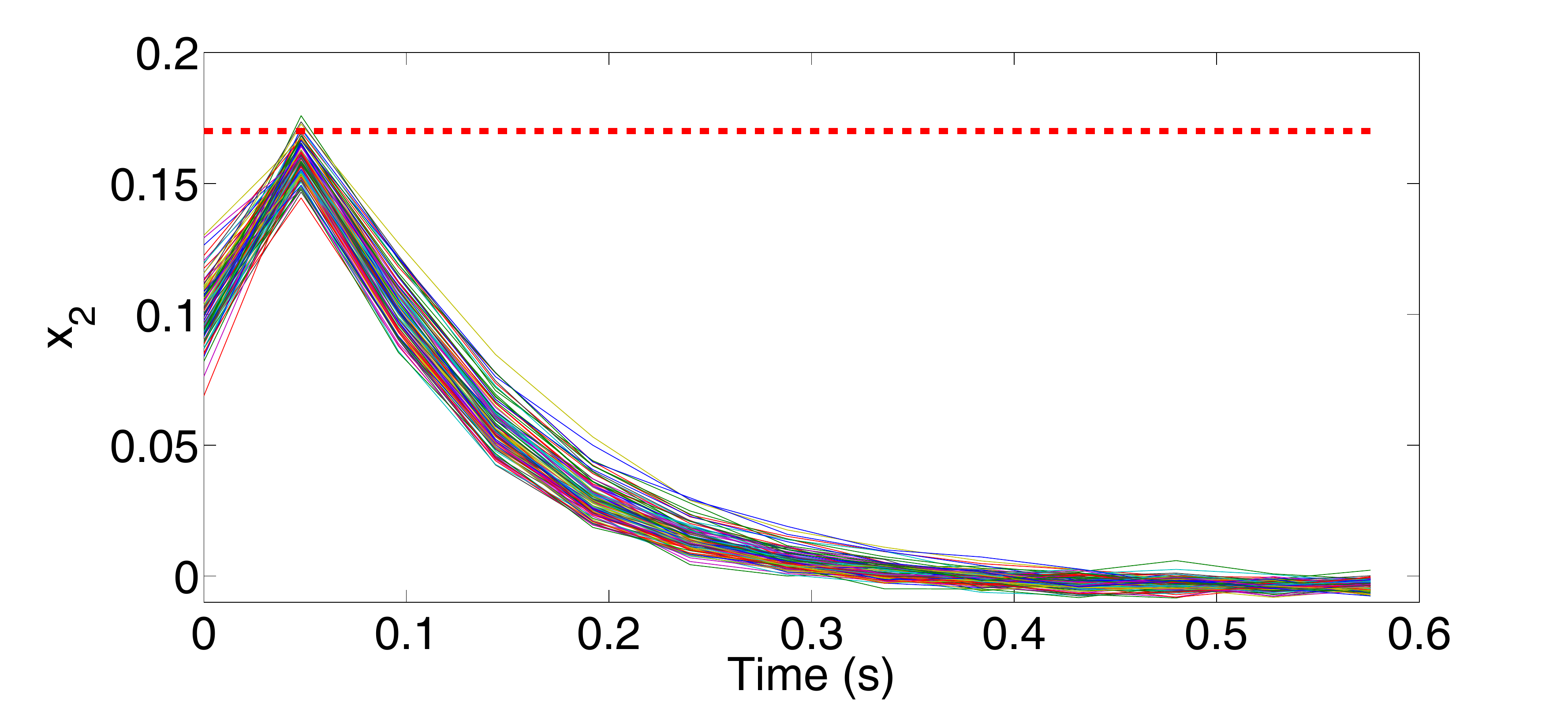}
}
\subfigure[Nominal MPC]{
\includegraphics[width=225pt]{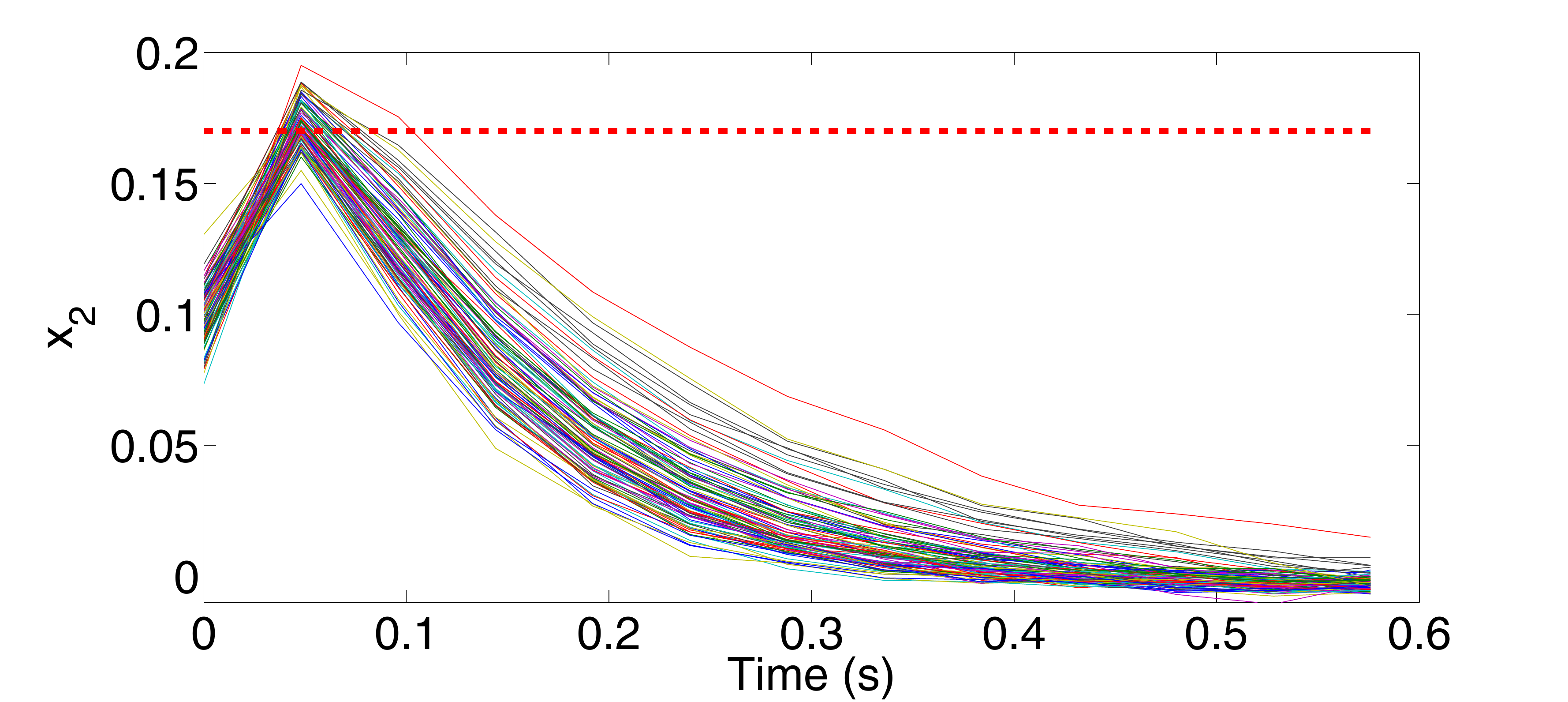}
}
\caption{Time profiles of $x_2$ obtained from $100$ closed-loop simulations of SMPC and nominal MPC. The red-dashed line represents the state constraint. Nominal MPC leads to constraint violation in $46\%$ of the cases.}
\label{fig2}
\end{figure}

The controller performance is evaluated based on $100$ closed-loop simulations in the presence of probabilistic uncertainties and process noise, and is compared to that of nominal MPC with terminal constraints. Fig.~\ref{fig1} shows the histograms of $x_1$ for both MPC approaches at three different times. SMPC leads to smaller mean (i.e., deviation with respect to the operating point) and smaller variance. This suggests that SMPC can effectively deal with the system uncertainties and process noise. Fig.~\ref{fig1} indicates that the state approaches the operating point ($x_1$ approaches zero). To assess the state chance constraint handling, the time profiles of $x_2$ for the $100$ runs are shown in Fig.~\ref{fig2}. The state constraint is fulfilled in over $95\%$ of simulations, whereas it is violated in nearly $46\%$ of closed-loop simulations of nominal MPC. Hence, the inclusion of the chance constraint into SMPC leads to effective state constraints satisfaction.            

\section{Conclusions}
\label{sec:Conclusions}

The paper presents a SMPC approach with state chance constraints for linear systems with time-invariant probabilistic uncertainties and additive Gaussian process noise. A tractable formulation for SMPC is presented. Closed-loop stability of SMPC is established for the unconstrained case.

\appendix

\subsection{Proof of Proposition 3}

From the definitions of the cost functions, $c(\mathbf{\Phi},\mathbf{K}\mathbf{\Phi}) = \| \mathbf{\Phi} \|^2_{\mathbf{Q} + \mathbf{K}^\top \mathbf{R} \mathbf{K}}$ and $\mathbf{E}\big[c_f\big((\mathbf{A}+\mathbf{B}\mathbf{K})\mathbf{\Phi}+\mathbf{F}w_0\big)| \mathbf{\Phi}\big] = \| (\mathbf{A}+\mathbf{B}\mathbf{K}) \mathbf{\Phi} \|^2_P + \text{tr}(\mathbf{F}^\top P \mathbf{F} \Sigma)$. Using the Lyapunov equation \eqref{eq:Lyap},
\begin{align*}
& c(\mathbf{\Phi},\mathbf{K}\mathbf{\Phi}) - c_f(\mathbf{\Phi})+ \mathbf{E}\big[c_f\big((\mathbf{A}+\mathbf{B}\mathbf{K})\mathbf{\Phi}+\mathbf{F}w_0\big) | \mathbf{\Phi}\big] = \\
& \mathbf{\Phi}^\top ((\mathbf{A}+\mathbf{B}\mathbf{K})^\top P (\mathbf{A}+\mathbf{B}\mathbf{K}) - P + \mathbf{M})\mathbf{\Phi} + \text{tr}(\mathbf{F}^\top P \mathbf{F} \Sigma) \\
& = -\delta \mathbf{\Phi}^\top \mathbf{M}\mathbf{\Phi} + \text{tr}(\mathbf{F}^\top P \mathbf{F} \Sigma).
\end{align*}
Note that $\delta \inf_{\mathbf{\Phi\in D}} (\mathbf{\Phi}^\top \mathbf{M}\mathbf{\Phi}) = 0$ such that the supremum of this expression is $\text{tr}(\mathbf{F}^\top P \mathbf{F} \Sigma) > 0$. Hence, there exists a number $b \geq 0$ that satisfies assertion \eqref{eq:Lem3_PhiinD}. For all $\mathbf{\Phi} \not\in D$, this expression will be less than or equal to zero such that assertion \eqref{eq:Lem3_PhinotinD} is also satisfied. $\hfill$ $\blacksquare$

\subsection{Proof of Proposition 4}

Define the $N$-length sequences $\tilde{L}^s := \{ K,\ldots,K\}$ and $\tilde{g}^s := \{ 0,\ldots,0 \}$. Let $\{ \mathbf{\Phi}^s_i \}_{i=1}^N$ denote the sequence obtained by applying the policy $(\tilde{L}^s,\tilde{g}^s)$ to the system \eqref{eq:Xdet} (i.e., $\mathbf{\Phi}^s_{i+1} = (\mathbf{A}+\mathbf{B}\mathbf{K})\mathbf{\Phi}^s_{i} + \mathbf{F}w_i,~ \forall i \in \mathbb{Z}_{[0,N-1]}$ with initial condition $\mathbf{\Phi}^s_{0} = \mathbf{\Phi}$ for any fixed $\mathbf{\Phi} \in \mathbb{R}^n$) and $V_N^s(\mathbf{\Phi}) := V_N(\mathbf{\Phi},\tilde{L}^s,\tilde{g}^s)$. From Proposition 3, it can be derived that
\begin{align*}
c_f(\mathbf{\Phi}) & \geq \mathbf{E}\big[c_f(\mathbf{\Phi}^f_{1}) | \mathbf{\Phi}^s_{0} = \mathbf{\Phi}\big] + c(\mathbf{\Phi},\mathbf{K}\mathbf{\Phi}) - b \\
c_f(\mathbf{\Phi}^s_1) & \geq \mathbf{E}\big[c_f(\mathbf{\Phi}^s_{2}) | \mathbf{\Phi}^s_{1}\big] + c(\mathbf{\Phi}_1^s,\mathbf{K}\mathbf{\Phi}_1^s) - b \\
& \vdots \\
c_f(\mathbf{\Phi}^s_{N-1}) & \geq \mathbf{E}\big[c_f(\mathbf{\Phi}^s_{N}) | \mathbf{\Phi}^s_{N-1}\big] + c(\mathbf{\Phi}_{N-1}^s,\mathbf{K}\mathbf{\Phi}_{N-1}^s) - b.
\end{align*}
Recursively substituting these expressions into one another yields $c_f(\mathbf{\Phi}) \geq V_N^s(\mathbf{\Phi}) - Nb$. Subtracting $V_N^\star(\mathbf{\Phi})$ from both sides of this inequality gives $c_f(\mathbf{\Phi}) - V_N^\star(\mathbf{\Phi}) \geq V_N^s(\mathbf{\Phi}) - V_N^\star(\mathbf{\Phi}) - Nb$. This assertion is due to $V_N^s(\mathbf{\Phi}) \geq V_N^\star(\mathbf{\Phi})$, since $(\tilde{L}^s,\tilde{g}^s)$ is a suboptimal policy to $(\tilde{L}^\star,\tilde{g}^\star)$ for arbitrary $\mathbf{\Phi} \in \mathbb{R}^n$. $\hfill$ $\blacksquare$

\subsection{Proof of Theorem 1}

From the last $N-1$ elements of $(\tilde{L}^\star,\tilde{g}^\star)$, define $\tilde{L}^f:=\{ L_1^\star,\ldots,L_{N-1}^\star,K \}$ and $\tilde{g}^f:=\{ g_1^\star,\ldots,g_{N-1}^\star,0 \}$ to be $N$-length feasible sequences of the control parameters. Let $V_N^f(\mathbf{\Phi}):=V_N(\mathbf{\Phi},\tilde{L}^f,\tilde{g}^f)$ and define $\mathbf{\Pi}^\star := \{ \mathbf{\Pi}^\star_0,\ldots,\mathbf{\Pi}^\star_{N-1} \}$ to be the optimal policy \eqref{eq:Udet}. Denote the ``optimal'' states $\{ \mathbf{\Phi}^\star_i \}_{i=0}^N$ (obtained by applying $\mathbf{\Pi}^\star$ to \eqref{eq:Xdet}) by 
\begin{align} \label{eq:Phistar}
\mathbf{\Phi}^\star_{i+1} = \mathbf{A} \mathbf{\Phi}^\star_i + \mathbf{B} \mathbf{\Pi}^\star_i(\mathbf{\Phi}^\star_i) + \mathbf{F} w_i,~\text{given} \; \mathbf{\Phi}^\star_0. 
\end{align}
From \eqref{eq:VN} and \eqref{eq:Phistar}, it is known that
\begin{align*}
& \mathbf{E}\big[ V_N^f(\mathbf{\Phi}^\star_{1}) | \mathbf{\Phi}^\star_0 = x \otimes e_{p+1} \big] - V_N^\star(x\otimes e_{p+1}) \\
& = \mathbf{E}\Big[ -\| \mathbf{\Phi}_0^\star \|^2_{\mathbf{Q}} - \| \mathbf{\Pi}^\star_0(\mathbf{\Phi}_0^\star) \|_\mathbf{R}^2 + \| \mathbf{\Phi}^\star_N \|^2_\mathbf{Q} + \| \mathbf{K}\mathbf{\Phi}^\star_N \|^2_\mathbf{R} \\
& + \| (\mathbf{A}+\mathbf{B}\mathbf{K})\mathbf{\Phi}^\star_{N} + \mathbf{F}w_N \|^2_P - \| \mathbf{\Phi}^\star_N \|^2_P \big| \mathbf{\Phi}^\star_0 = x\otimes e_{p+1} \Big].
\end{align*}
The first two terms above can be taken out of the expected value, and can be derived to be $\| x\otimes e_{p+1} \|^2_{\mathbf{Q}} + \| \mathbf{\Pi}^\star_0(x\otimes e_{p+1}) \|_\mathbf{R}^2 = \| x \|^2_Q + \| \mathbf{\pi}^\star_0(x) \|^2_R$. The law of iterated expectation for Markov processes and Proposition 3 are used to obtain a bound on this expression
\begin{align*}
& \mathbf{E}\bigg[ \mathbf{E}\Big\{ \| \mathbf{\Phi}^\star_N \|^2_\mathbf{Q} + \| \mathbf{K}\mathbf{\Phi}^\star_N \|^2_\mathbf{R} + \| (\mathbf{A}+\mathbf{B}\mathbf{K})\mathbf{\Phi}^\star_{N} + \mathbf{F}w_N \|^2_P \\[-2mm]
& ~~~~~~~~ - \| \mathbf{\Phi}^\star_N \|^2_P \big| \{ \mathbf{\Phi}^\star_{s} \}_{s=0}^N \Big\} \big| \mathbf{\Phi}^\star_0 = x\otimes e_{p+1} \bigg] \\
& \leq \mathbf{E}\big[ b \mathcal{I}_{D}(\mathbf{\Phi}_N^\star) \big| \mathbf{\Phi}^\star_0 = x\otimes e_{p+1} \big] \\
& = b \mathbf{Pr}\big[ \mathbf{\Phi}_N^\star \in D \big| \mathbf{\Phi}^\star_0 = x\otimes e_{p+1} \big] \leq b.
\end{align*}
This result is applied to the starting expression to derive $\mathbf{E}\big[ V_N^f(\mathbf{\Phi}^\star_{1}) | \mathbf{\Phi}^\star_0 = x \otimes e_{p+1} \big] - V_N^\star(x\otimes e_{p+1}) \leq -( \| x \|^2_Q + \| \mathbf{\pi}^\star_0(x) \|^2_R ) + b$. From the optimality of $(\tilde{L}^\star,\tilde{g}^\star)$ and from assumption \eqref{eq:assumption}, it is known that
\begin{align*}
& \mathbf{E}\big[ V_N^\star(x_1 \otimes e_{p+1}) | x_0 = x  \big] - V_N^\star(x\otimes e_{p+1}) \\
& \leq \mathbf{E}\big[ V_N^f(x_1 \otimes e_{p+1}) | x_0 = x  \big] - V_N^\star(x\otimes e_{p+1}) \\
& \leq \mathbf{E}\big[ V_N^f(\mathbf{\Phi}^\star_{1}) | \mathbf{\Phi}_0 = x \otimes e_{p+1} \big] - V_N^\star(x\otimes e_{p+1}) \\
& \leq -\big( \| x \|^2_Q + \| \mathbf{\pi}^\star_0(x) \|^2_R \big) + b \leq -\| x \|^2_Q + b.
\end{align*}
For some constant $\alpha \in [0,1)$, define the set $D' := \{ z\in \mathbb{R}^{n_x} | z^\top Q z \leq \alpha (z \otimes e_{p+1})^\top \mathbf{P} (z \otimes e_{p+1})\}$ such that $\mathbf{E}\big[ V_N^\star(x_1 \otimes e_{p+1}) | x_0 = x  \big] - V_N^\star(x\otimes e_{p+1}) \leq -\alpha c_f(x\otimes e_{p+1}) + b,~ \forall x \not\in D'$. From Proposition 4, $-\alpha c_f(x \otimes e_{p+1}) \leq -\alpha V_N^\star(x \otimes e_{p+1}) + \alpha Nb,~ \forall x \in \mathbb{R}^{n_x}$ such that $\mathbf{E}\big[ V_N^\star(x_1 \otimes e_{p+1}) | x_0 = x  \big] - V_N^\star(x\otimes e_{p+1}) \leq -\alpha V_N^\star(x \otimes e_{p+1}) + b(1+\alpha N),~ \forall x \not\in D'$. Since $\lim_{\| z \|\rightarrow +\infty} c(z \otimes e_{p+1},\pi^\star_0(z)\otimes e_{p+1}) = +\infty$, \eqref{eq:VN} implies that $\lim_{\| z \|\rightarrow +\infty} V_N^\star(z\otimes e_{p+1}) = +\infty$. Hence, there must exist a closed ball $D''$ around the origin $0 \in \mathbb{R}^{n_x}$ of a radius large enough such that $V_N^\star(z\otimes e_{p+1}) \geq 2b (\alpha^{-1} + N),~\forall z \not\in D''$ \cite{cha13}. Substituting this into the previous expression leads to
\begin{align*}
& \mathbf{E}\big[ V_N^\star(x_1 \otimes e_{p+1}) | x_0 = x  \big] - V_N^\star(x\otimes e_{p+1}) \\
& ~~~~~~ \leq -\frac{\alpha}{2} V_N^\star(x \otimes e_{p+1}),~~\forall x \not \in D''.
\end{align*}
The sets $D'$ and $D''$ should satisfy $D' \subseteq D'' \subset \mathbb{R}^{n_x}$ such that $x \not\in D'' \Rightarrow x \not\in D'$. This represents a geometric drift condition outside the compact set $D''$ such that the assertion follows directly from Proposition 2. $\hfill$ $\blacksquare$

%
%

\bibliographystyle{ieeetr}
\bibliography{Literature_list}

\end{document}